\documentclass{elsart}
\usepackage{graphicx}
\usepackage{bbm}

\makeatletter

\def\vereq#1#2{\lower3pt\vbox{\baselineskip1.5pt \lineskip1.5pt
\ialign{$\m@th#1\hfill##\hfil$\crcr#2\crcr\sim\crcr}}}

\makeatother

\begin{document}
\runauthor{Hencken, Baur, Trautmann}
\begin{frontmatter}
\title{A cluster version of the GGT sum rule}
\author[UBA]{Kai Hencken}
\author[FZJ]{Gerhard Baur}
\author[UBA]{Dirk Trautmann}
\address[UBA]{Institut f\"ur Physik, Universit\"at Basel,
4056 Basel, Switzerland}
\address[FZJ]{Forschungszentrum J\"ulich, 52425 J\"ulich, Germany}
\maketitle

\begin{abstract}
We discuss the derivation of a ``cluster sum rule'' from the 
Gellmann-Goldberger-Thirring (GGT) sum rule as an alternative to 
the Thomas-Reiche-Kuhn (TRK) sum rule, which was used as the basis up to 
now. We compare differences in the assumptions and approximations. Some
applications of the sum rule for halo nuclei, as well as, nuclei with a 
pronounced cluster structure are discussed.
\end{abstract}
\end{frontmatter}
%pacs: 25.20.-x, 21.60.Gx
\section{Introduction}
Sum rules like the GGT and the TRK sum rule,
have played an important role in the understanding of global properties 
of (well) bound system and their excitation spectrum. One of their major 
features is their universality, as they cannot only be applied to nuclei 
but also, e.g., to atoms or the substructure spectrum of the nucleus. 
For the derivation however one makes a number of assumptions; deviations 
of the experimental findings from the pure results are expected and found. 
Both the classical TRK and GGT sum rule were found experimentally 
for nuclei to overfulfil the theoretical result up to a factor 
of two.

More recently the TRK sum rules has been applied to the domain of
very proton and very neutron rich nuclei or systems with a pronounced
cluster structure. For these systems a so-called {\em ``cluster sum rule''} 
was derived \cite{ABG82} and is has lead to some insight in the low-lying 
dipole strength (sometimes called the {\em`` pigmy resonance''}) in such 
systems
\cite{Ikeda92}. We want to show here that this cluster sum rule can be derived 
also from the Gellmann-Goldberger-Thirring (GGT) sum rule. Whereas this 
approach leads ultimately to the same mathematical expression as the TRK 
cluster sum rule (as it does already for the usual ``non-cluster'' sum rule), 
the approximations and assumptions are quite different here and therefore some 
further insight into the nature of the cluster sum rule can be gained. One 
advantage of the GGT cluster sum rule is, that it is not based on the 
long-wavelength limit and Siegert's theorem and that it is not restricted to 
the dipole excitation spectrum alone. We review briefly the foundations of the
TRK cluster sum-rule and the approximations used before deriving the GGT 
version. We apply the result then to systems with a halo-structure and with 
two or more clusters. 

\section{Review of the TRK cluster sum rule}

The classical TRK sum rule starts from the dipole operator strength for the 
transition of a nucleus from the ground state $i$ to some excited state 
$f$
\begin{equation}
f_{fi} = \frac{2 m_N e^2}{\hbar^2} 
(E_f - E_i) \left|\left< f | \vec D | i\right>\right|^2.
\end{equation}
with $\vec D$ the dipole operator $\frac{NZ}{A}
\left( \vec R_p - \vec R_n\right)$, where $\vec R_p$ is the c.m. position of 
all protons $\frac{1}{Z} \sum_p \vec r_p$ and $\vec R_n$ of all neutrons
$\frac{1}{N} \sum_n \vec r_n$ respectively.
The dipole operator $\vec D$ is written in a way that the center of mass
motion of the whole system is taken out.
In order to sum this equation over all excited states $f$, one replaces $E_f$
and $E_i$ with the Hamilton operator operating either on $|i>$ or $|f>$ and 
makes use of closure to get
\begin{equation}
\sum_f f_{fi} = \frac{m_N e^2}{\hbar^2} 
\left<i | \left[\vec D,\left[H,\vec D\right]\right]|i\right>.
\label{eq:summed}
\end{equation}
For a Hamiltonian, which consists of a nonrelativistic kinetic term together 
with a local (momentum independent) potential $V$, the double commutator 
in Eq.~(\ref{eq:summed}) is found to be
$\hbar^2/m_N$. Using the relation between the dipole strength $f_{fi}$ and 
the total photoabsorption cross section $\sigma_{\gamma}$ based on Siegert's
theorem (and therefore on the long wavelength approximation)
one gets the TRK sum rule
\begin{equation}
\int_0^\infty 
\sigma_\gamma(\omega) d\omega = \frac{2\pi^2 \hbar e^2}{m_N c} \frac{N Z}{A}
= 60 \frac{N Z}{A} \mbox{MeV mb}.
\end{equation}
The important feature of this sum rule is the fact, that it only depends on
the number of protons and neutrons and is completely independent of their
arrangement within the nucleus.

Recently the same idea has been used in the case of a system composed of
two clusters of nuclei with charge $Z_a$ and $Z_b$ ($Z=Z_a+Z_b$) and mass 
number $A_a$ and $A_b$ ($A=A_a+A_b$) \cite{ABG82,Ikeda92,Bertsch98}. The 
idea is a decomposition of 
the dipole operator $\vec D$ into ``external'' and ``internal'' coordinates. 
Using
\begin{equation}
\vec D = \vec D_a + \vec D_b + \vec D_{(ab)}
\label{eq:ddef}
\end{equation}
where $\vec D_a$ and $\vec D_b$ are the dipole moments of each cluster $a$ and
$b$
\begin{equation}
\vec D_i = \sum_{p\in i} (\vec r_p - \vec R_i) 
\label{eq:didef}
\end{equation}
for $i=a,b$. The sum in Eq.~(\ref{eq:didef}) 
goes only over those protons being part of cluster $i$.
$\vec D_{(ab)}$ is the relative dipole moment between $a$ and $b$ 
\begin{equation}
D_{(ab)} = Z_a (\vec R_a - \vec R) + Z_b (\vec R_b - \vec R)
 = e \left[ (Z_a A_b - Z_b A_a)/A \right] \vec S.
\end{equation}
$\vec R$ is the position of the center of mass of the nucleus, 
$R_a$ and $R_b$ are the position of the centers of mass of $a$ and $b$.
Finally
$\vec S$ denotes the vector connecting the center of mass of $a$ and $b$,
$\vec S = \vec R_a - \vec R_b$. Using this decomposition and looking at the
part of the photo absorption cross section coming from $\vec D_{(ab)}$ 
one gets a ``cluster version of the TRK sum rule''
\begin{equation}
\int_0^\infty \sigma_{(ab)}(\omega)d\omega = 
\frac{2\pi^2 \hbar e^2}{m_N c} \frac{(Z_a A_b - Z_b A_a)^2}{A A_a A_b}.
\end{equation}
where the cross section $\sigma_{(ab)}(\omega)$ is connected with
the relative dipole excitation due to $D_{(ab)}$.
Let us emphasize at this point, that the derivation of this cluster sum rule
makes no assumption, whether the decomposition of the system into two clusters
$a$ and $b$ is also expected to be of a physical nature, that is, whether the
nucleons in the nucleus are expected to be configured mainly into two 
subsystems $a$ and $b$.

If one on the other hand is looking at the specific breakup channel, where 
only the ``external'' coordinates $\vec R_a$ and $\vec R_b$ are important, 
that is, the channel $(ab) \rightarrow a+b$, the integral
\begin{equation}
\int_0^\infty \sigma_{(\gamma+(ab) \rightarrow a+b)} d\omega
\approx \frac{2\pi^2 \hbar e^2}{m_N c} 
\frac{(Z_a A_b - Z_b A_a)^2}{A A_a A_b}.
\end{equation}
is a measure of how much a supposed cluster configuration contributes to
the dipole sum, see also the comments in \cite{ABG82}.
In this integral one assumes that the excitation to bound cluster 
configurations of $a$ and $b$ do not contribute considerably to the integral.

This sum rule plays an important role in connection with the question 
of the existence of low lying strength in exotic nuclei. If one assumes that 
at low excitation energies it is preferable that the two clusters (which are 
assumed to be loosely bound with respect to each other
as compared to the internal binding) move against each 
other, the cluster sum rule gives a value for the strength, that is expected
to lie at low energies \cite{Ikeda92,BaurHT03}. Measurements especially of the 
neutron rich He isotopes and also O isotopes have confirmed the
existence of this low lying strength, which is often also called the 
{\em ``pigmy resonance''} in analogy to the giant dipole resonance, where most
of the strength of the TRK sum rule is concentrated.

Let us review the assumptions that went into the derivation of this sum rule:
It is based on the dipole operator $\vec D$ alone, therefore other multipole 
apart from the $E1$ transition are assumed to be small. It makes use of a 
Hamiltonian with a local potential, contributions coming especially from 
exchange currents are assumed to be small. In order to relate the dipole 
operator to the total photoabsorption cross section Siegert's theorem is 
used, therefore the sum rule is expected to be valid only at low excitation 
energies, where the long wavelength limit is valid.

Another problem, even though not relevant in the cases discussed here, is the 
nonrelativistic nature of the derivation. There is
no known derivation in the relativistic case, that is, the TRK sum rule is a 
nonrelativistic sum rule with no relativistic analogon. Some of these problems
are not present in the GGT sum rule, therefore we want to see whether the 
cluster sum rule can also be found in this approach.

\section{Derivation of the cluster sum rule from the GGT sum rule}

A different sum rule is the GGT sum rule \cite{EG88,deShalitF,Weise74};
this sum rule only makes assumptions about causality and 
analyticity of the forward elastic scattering amplitude $f(\omega)$
and uses a dispersion relation to relate the real and imaginary part of 
this amplitude. One gets the once subtracted dispersion relation \cite{EG88}
\begin{equation}
Re f(\omega_0) - Re f(0) = \frac{2\omega_0^2}{\pi} P.P. \int_0^\infty
d\omega \frac{Im f(\omega)}{\omega(\omega^2-\omega_0^2)}.
\end{equation}
Making use of the optical theorem, the imaginary part of the forward elastic
scattering amplitude is related to the total cross section $\sigma(\omega)$ 
by
\begin{equation}
\sigma(\omega) = \frac{4\pi \hbar c}{\omega} Im f(\omega).
\end{equation}
In addition the forward elastic scattering amplitude at $\omega=0$ is real and 
given by the Thomson limit
\begin{equation}
f(0) = Re f(0) = - \frac{(Z e)^2}{M c^2},
\end{equation}
where $Ze$ is the total charge and $M=m_N A$ the total mass of the system.
Let us assume that we have a system $c$ composed of two 
subsystems (``clusters'') $a$ and $b$. The more general case with more than 
two subsystems will be discussed below. In the limit 
$\omega_0 \rightarrow \infty$ one gets a relation for $i=a,b,c$
\begin{equation}
Re f_i(\infty) + \frac{Z_i^2 e^2}{A_i m_N c^2} = - \frac{1}{2\pi^2 \hbar c} 
\int_0^\infty d\omega \ \sigma_i(\omega).
\end{equation}

This relation is the basis of the usual GGT sum rule and also of a more
general cluster sum rule. A first assumption is that the scattering 
amplitude $f(\omega)$ at infinity of the system $c$ is just given by 
the sum of the scattering cross section of the two components $a$ and $b$, 
that is,
\begin{equation}
Re f_c(\infty)-Re f_a(\infty)-Re f_b(\infty)\approx 0
\end{equation}
(which is not true strictly speaking due to hadronic components in the photon 
and shadowing corrections in the nucleus, see below)
and therefore
\begin{eqnarray}
\int_0^\infty d\omega \left[ \sigma_c(\omega)-\sigma_a(\omega)-\sigma_b(\omega)
\right]&=& \frac{2\pi^2 \hbar e^2 }{m_N c} \left[
\frac{Z_a^2}{A_a}+\frac{Z_b^2}{A_b}-\frac{Z_c^2}{A_c}
\right]\label{eq:ggtsr}
\\
&=& \frac{(Z_a A_b - Z_b A_a)^2}{A_a A_b A_c}
\mbox{60 MeV mb},
\end{eqnarray}
which is the main result. For the more general case of
the decomposition of our system $c$ into $N$ different cluster the result
is
\begin{eqnarray}
\int_0^\infty d\omega \left( \sigma_c(\omega)-\sum_{i=1}^N \sigma_i(\omega)
\right)
&=& \frac{2\pi^2 \hbar e^2}{m_N c} \left[
\sum_{i=1}^N\frac{Z_i^2}{A_i}-\frac{Z_c^2}{A_c} \right]
\mbox{MeV mb}.
\label{eq:ggtN}
\end{eqnarray}

The usual GGT sum rule can be recovered from this expression by using
as cluster each nucleon, that is $Z=Z_c$ protons and $N=N_c$ neutrons.
With this we get the usual form of the GGT sum rule
\begin{eqnarray}
\int_0^\infty d\omega \left[ \sigma(\omega)-Z\sigma_p(\omega)-N\sigma_n(\omega)
\right]
&=& \frac{2\pi^2 \hbar e^2 }{m_N c} 
\frac{N Z}{A},
\end{eqnarray}
which coincides with the expression for the TRK sum rule, as already
did the expression for the cluster sum rule.

Let us decompose the total photoabsorption cross sections in 
Eq.~(\ref{eq:ggtsr}) according to the final state the nucleus $c$ goes to.
We can write the cross section as
\begin{equation}
\sigma_c = \sigma_{c +\gamma \rightarrow a+b} + 
\sigma_{c+\gamma\rightarrow a + X_b} +
\sigma_{c+\gamma\rightarrow X_a + b} +
\sigma_{c+\gamma\rightarrow X_a + X_b}+
\sigma_{c+\gamma\rightarrow c^*},
\end{equation}
where $X_a$ and $X_b$ denote ``fragments'' of $a$ and $b$, that is, all final
states not including $a$ or $b$ and $c^*$ denotes excited bound states of $c$.
Assuming that due to the clustering structure the breakup of $a$ in the 
nucleus $c$ is the same as the one of $a$ alone (and a shift in energy due to 
the binding of $a$ is not important in the integration), we have 
({\em ``spectator approximation''}) 
\begin{equation}
\sigma_{c+\gamma\rightarrow a + X_b} \approx
\sigma_{b+\gamma\rightarrow X_b},\quad
\sigma_{c+\gamma\rightarrow X_a + b} \approx
\sigma_{a+\gamma\rightarrow X_a},\quad
\sigma_{c+\gamma\rightarrow X_a + X_b} \approx 0.
\end{equation}
Also for the loosely bound system we are mainly looking at there is only
a small number of excited states and we have
\begin{equation}
\sigma_{c+\gamma\rightarrow c^*} \approx 0
\end{equation}
(this approximation is not necessary, as $\sigma_{c+\gamma\rightarrow c^*}$
can easily be included on the left hand side of Eq.~(\ref{eq:ggtsr2}))
and find within the ``spectator limit'' of the GGT cluster sum rule
\begin{equation}
\int_0^\infty d\omega \sigma_{c+\gamma\rightarrow a+b}(\omega)
\approx 
\frac{(Z_a A_b - Z_b A_a)^2}{A_a A_b A_c}
\mbox{60 MeV mb},
\label{eq:ggtsr2}
\end{equation}
a relation, which can be measured in experiments and used as a test for
the contribution of the clustering component to the cross section.

As was already the case for the GGT sum rule, where the excitation of 
the individual nucleons occur only above the pion production threshold, also 
here the excitation at lower energies is dominated by the relative
excitation of $a$ and $b$, especially if the subsystems $a$ and $b$ are well 
bound systems like $\alpha$ or even nucleons.
But whereas the pion production threshold is very high compared to typical
nuclear excitation energies, this is not the case here most of the time. Still
the sum rules of Eq.~(\ref{eq:ggtsr}) or Eq.~(\ref{eq:ggtsr2}) integrated up
to infinity are valid.

In the case of more than two clusters the situation is more complex. Assuming
that our system $c$ consists of three clusters $a$, $b$, $d$, we have the added
complication that the system can have bound states of $a$ and $b$ or any other
binary system. In this case it is easy to show that the difference of the cross
sections in the GGT cluster sum rule Eq.~(\ref{eq:ggtN}) in the spectator 
approximation corresponds to all channels with final states
composed of bound or continuum states of the components $a$, $b$ and $d$. 
This will be relevant for the study of $^8$He below.

As in the TRK cluster sum rule no assumption about the validity of the 
clustering in the nuclear structure of $c$ was made in the 
derivation of Eq.~(\ref{eq:ggtsr}). On the other hand the last form in 
Eq.~(\ref{eq:ggtsr2}), based on the spectator approximation, 
of course strongly assumes that the major contribution of the excitation
cross section comes from the excitation of the relative motion of $a$ and $b$
and therefore is only true if the system is dominated by this clustering 
structure. Deviations of the second form Eq.~(\ref{eq:ggtsr2}) can therefore 
be seen as a test of the clustering hypothesis.

As in the TRK sum rule also in the GGT sum rule there are deviations
from the simple picture. Especially the assumption, that the
difference $Re f_c(\infty)-Re f_a(\infty)-Re f_b(\infty)$ is zero is
not really true. At high energies the hadronic component of the photon
mainly interacts with the nucleons, making the nucleons black objects
and therefore the nucleons are shadowing each other. This shadowing
effects leads to an enhancement of up to a factor of two compared to the
theoretical result. This is discussed extensively in \cite{Weise73a,Weise73b}.
On the other hand one expects that the deviation of the experimental
results from the theoretical prediction of the GGT cluster sum rule is smaller
as for the GGT sum rule itself. If the average distance
between the two clusters $a$ and $b$ is large (ideally larger than the size 
of each cluster itself), shadowing corrections 
of $a$ on $b$ or $b$ on $a$ will be small, as the
two clusters $a$ and $b$ do not block each other very much and shadowing 
within the clusters $a$ or $b$ do not lead to deviations.

Let us finally review the advantage of this derivation of the cluster
sum rule: It is not based on any specific model of the system but only
on general properties, like causality, of the forward
elastic scattering amplitude. It is
therefore valid for all multipole moments, not only for the dipole moment, 
and it is independent of the validity of the long wavelength limit, that is the
validity of the Siegert's theorem. It is also a sum rule which is the same
in the nonrelativistic as well as in the relativistic case.
This is also the reason why the TRK sum rule is often found to hold also in
relativistic models \cite{CohenL98}.

\section{Application to nuclei}
The TRK cluster sum rule has been applied already in some cases, e.g., for
$^{11}$Be, \cite{HansenJ87,Ikeda92}, the neutron rich helium isotopes $^6$He
and $^8$He \cite{Aumann99,Meister02,Markenroth01} and also for neutron rich 
oxygen isotopes \cite{Leistenschneider01}.
We want to show how the analysis made in these cases can be taken further by 
using the more generalised versions of the cluster sum rule. In addition we 
will also look at cluster nuclei, especially $^{6}$Li, $^{16}$O and $^{9}$Be.

A detailed analysis of the electromagnetic dissociation of both $^6$He and 
$^8$He was made at GSI \cite{Meister02}. Here the fragments after the 
Coulomb breakup reaction $^6\mbox{He} \rightarrow {}^4\mbox{He} + 2 \mbox{n}$ 
and $^8\mbox{He} \rightarrow {}^6\mbox{He} + 2 \mbox{n}$ 
where measured and from the invariant mass 
the excitation energy, that is, the photon energy was reconstructed. The cross 
section integrated over a range of photon energies (mainly limited by the
experimental setup) was also calculated and compared with the prediction 
coming from the cluster sum rule under the assumption of a
$^4$He cluster together with the ``halo-neutrons'' making up the other 
cluster. Based on this analysis it was found that the cluster sum rule is 
almost exhausted in the case of $^6$He. 

We can find cluster sum rules also for other configurations. Therefore we
want to reevaluate the findings especially of \cite{Meister02} in this light.
\begin{table}[tbp]
\caption{The predictions of the GGT cluster sum rule are shown for different
configurations of the cluster for $^6$He and $^8$He. The cross section 
differences on which the sum rule is based are shown together with the 
channels that this difference corresponds to in the limit of the spectator
approximation. The last column gives the experimental results of 
\protect\cite{Meister02}.}
\label{tab:he}
\begin{tabular}{c|c|c|c}
$\sigma$ difference & ``spectator approx.''&
 GGT result & exp. result\\
  &  & (MeV mb) & (MeV mb)\\
\hline
$\sigma(^8\mbox{He})-\sigma(^6\mbox{He})-2\sigma(n)$
& $\sigma(\gamma + {}^8$He $\rightarrow$ $^6$He + 2n) 
& $1/6 \times 60 = 10$ & $7.5 \pm 1.4$\\
\hline
$\sigma(^8\mbox{He})-\sigma(^4\mbox{He})-4\sigma(n)$
&
\begin{tabular}{c}
$\sigma(\gamma + {}^8$He $\rightarrow$ $^6$He + 2n) +\\
$\sigma(\gamma + {}^8$He $\rightarrow$ $^4$He + 4n)
\end{tabular}
& $1/2 \times 60 = 30$ & ---\\
\hline
$\sigma(^6\mbox{He})-\sigma(^4\mbox{He})-2\sigma(n)$
&$\sigma(\gamma + {}^6$He $\rightarrow$ $^4$He + 2n )
& $1/3 \times 60 = 20$ & $26 \pm 5$\\
\end{tabular}
\end{table}

From Table~\ref{tab:he} one can see, that within the spectator approximation
of the GGT sum rule
1/3 of the dissociation cross section of $^8$He to a final state including
an $^4$He nucleus goes into the channel $^6$He + 2n, 2/3 is expected to go 
directly to $^4$He+ 4n, the channel, which was not
measured in the experiment, see \cite{Meister02}. On the other hand it was 
found that the dissociation cross section to $^6$He+2n (in the 
energy range measured in the experiment, that is below 10 MeV) is about 
1/3 of the one for the dissociation of $^6$He to $^4$He + 2n and exhausts 
the cluster sum rule based on a $^4$He cluster and four neutrons to about 
25\%. This is close to the expected 33\% from the analysis above. One 
expects a 
large cross section for the channel going to $^4$He and four neutrons directly.
In \cite{Aumann99} it was already found that for $^6$He to $^4$He + 2n the
cluster sum rule is exhausted to almost 100\%.

As the cross section for a specific channel was measured in this case, this
is a test of the validity of the spectator approximation and therefore a test
of the cluster configuration of $^8$He and $^6$He. The
experimental results are in agreement with the hypothesis that $^8$He is 
predominantly in a structure with two neutrons building one cluster and 
$^6$He the second.
This doesn't mean that $^8$He is a two-neutron halo system build around $^6$He,
but does indicate that two neutrons form a system that is more or less 
decoupled from the rest of the system. A configuration with a $^4$He core
together with two 2n cluster would fulfil the sum rule as well.

A second application can be made in the case of the neutron rich oxygen 
isotopes as measured in \cite{Markenroth01}. The electromagnetic dissociation 
was studied for $A=17,19,20,21,22$ and it was found that for large neutron 
excess, that is large $A$, there is a tendency for the appearance of low lying
dipole strength. The experiment measured only the cross section for the
electromagnetic breakup with the emission of up to three neutrons, assuming
that due to the large threshold for proton emission, this 
is to a good approximation identical to the total photoabsorption cross 
section. The authors of \cite{Markenroth01} only made a comparison with the 
prediction of the cluster sum rule based on the 
assumption of one cluster being the $^{16}$O-core whereas all other
neutrons are part of the second ``cluster''. 
Only the sum of $x$n neutron emission cross sections was published and no 
individual data for 1n, 2n or 3n. 

We are analyzing their data under the assumption that the emission
of up to three neutrons is to a good approximation already the full 
photoabsorption cross section, as was done in 
\cite{Leistenschneider01}. In Table~\ref{tab:oxy} below we give the differences
of the integrated cross sections extracted from Fig.~3 of 
\cite{Leistenschneider01} together with the sum rule prediction based on the
differences of the integrated cross section for the $A$ and $A-1$, $A-2$ or 
$A-3$ isotopes.
\begin{table}[tbp]
  \caption{An analysis of the difference of the photoabsorption cross sections
for $^A$O and $^{(A-1)}$O, $^{(A-2)}$O and $^{(A-3)}$O 
from the GGT cluster sum rule. The theoretical expected values are compared
with experimental results from \protect\cite{Leistenschneider01}. Please
note that the experimental results include only photon energies up to 20~MeV
and reactions with up to 3 neutrons in the final state. Also shown is a
comparison of the results by applying the spectator model, assuming that
$^A$O consists of a $^{(A-3)}O$ and a $3n$ cluster.
The last column shows the threshold values for the reactions. 
See the text for details of the analysis.}
\label{tab:oxy}
  \centering
  \begin{tabular}{c|c|c|c}
  cross section diff. & GGT result & exp. result & 
threshold \\
                      & (MeV~mb) & (MeV~mb) & (MeV)\\
\hline
  $\sigma(^{22}$O)-$\sigma(^{21}$O) & 8.34 & 10.1  & 6.849\\
  $\sigma(^{21}$O)-$\sigma(^{20}$O) & 9.12 & -22.7 & 3.807\\
  $\sigma(^{20}$O)-$\sigma(^{19}$O) & 10.1 & 19.1  & 7.608\\
  $\sigma(^{19}$O)-$\sigma(^{18}$O) & 11.3 &  7.1  & 3.957\\
  $\sigma(^{18}$O)-$\sigma(^{17}$O) & 12.5 &  31   & 8.044\\
\hline
  $\sigma(^{22}$O)-$\sigma(^{20}$O) & 17.5  & -12.6&10.656\\
  $\sigma(^{21}$O)-$\sigma(^{19}$O) & 19.2 & -3.6  &11.415\\
  $\sigma(^{20}$O)-$\sigma(^{18}$O) & 21.4 & 26.2  &11.565\\
  $\sigma(^{19}$O)-$\sigma(^{17}$O) & 23.8 & 38.1  &12.001\\
\hline
  $\sigma(^{22}$O)-$\sigma(^{19}$O) & 27.5 & 6.5   &18.264\\
  $\sigma(^{21}$O)-$\sigma(^{18}$O) & 30.5 & 3.5   &15.372\\
  $\sigma(^{20}$O)-$\sigma(^{17}$O) & 33.9 & 57.2  &19.609\\
\hline
 spectator model & & & \\
\hline
  $\sigma(^{22}$O$\rightarrow {}^{19}$O + 3n$)$ & 27.5 & 62.1 & \\
  $\sigma(^{21}$O$\rightarrow {}^{18}$O + 3n$)$ & 30.5 & 52.7& \\
  $\sigma(^{20}$O$\rightarrow {}^{17}$O + 3n$)$ & 33.9 & 74.9& \\
  $\sigma(^{19}$O$\rightarrow {}^{16}$O + 3n$)$ & 38.9 & 55.7& 
  \end{tabular}
\end{table}

Please recall that by taking the difference of two integrated cross sections,
we are comparing the experimental results with the prediction of 
Eq.~(\ref{eq:ggtsr}), the cluster sum rule using the difference of the 
photoabsorption cross section. This sum rule doesn't use the spectator 
approximation and therefore should be fulfilled independent of the fact,
whether the $O$ isotope is clustered into a core and either one, two or
three decoupled neutrons. The agreement with the theoretical expectation
for the difference of the integrated cross sections for $A$ and 
$A-1$ and $A-2$ is not too bad in some cases, but the agreement of the 
difference between $A$ and $A-3$ is not good. The discrepancy can 
mainly be 
attributed to the limitation of the experimental results: The sum rule
is calculated only for energies up to 20~MeV (whereas the data seem to be 
available up to 30~MeV) and therefore some contributions at higher energies
are missing. Also mentioned in the table is the threshold for the emission of
one, two or three neutrons.
Especially for the difference $\sigma(A)-\sigma(A-3)$ the fact 
that only the emission of up to three neutrons was measured is clearly 
important in order to compare the difference of the total photoabsorption
cross section. 
As the emission of up to three neutrons was measured, we can also test the
cluster hypothesis based on the spectator approximation and 
Eq.~(\ref{eq:ggtsr2}). This is a test of the (unrealistic) assumption that the 
oxygen isotope consist of a $A-3$-core together with three decoupled neutrons.
The results of this are shown in the last rows of the table, where we take
the experimental data as the cross section to be compared with the
theoretical result. As expected the agreement is not good, stating that in
none of the cases (with maybe the possible exception of $^{19}O$, which can be
thought to consist of the $^{16}O$ core and three neutrons) we expect
a three neutron cluster. A comparison of the GGT cluster sum rule prediction 
with the results of the reaction channel 
$\sigma(\gamma+A \rightarrow (A-1) + n)$ and 
$\sigma(\gamma+A \rightarrow (A-2) + 2n)$
would be more interesting.
These results, even though measured in \cite{Leistenschneider01}
are not quoted by the authors.

Finally let us also look at systems, which consist of more complex structures
than the halo nuclei: Photodissociation of $^6$Li and $^{16}$O 
at low photon energies are of astrophysical interest \cite{BaurHT03} 
and several attempts have been 
made to determine these cross sections accurately. In the usual TRK approach 
only the dipole transition is taken into account and it is found that no dipole
transition exists in the breakup of a system into two clusters with the same 
$Z/A$-ratio (as the effective charge is zero). Here we find that this result 
is also true in the GGT approach, where no assumption about the multipolarity 
of the absorbed photon is made. Therefore any contribution to the total 
integral of the cross section at low energies must either be compensated by a 
difference in the cross section at higher energies or must come from deviations
of the sum rule.
For the reactions $^6\mbox{Li} + \gamma \rightarrow \alpha + d$ and
$^{16}\mbox{O} + \gamma \rightarrow ^{12}C + \alpha$ the spectator 
approximation also predicts a zero result. Here of course the cross section
then mainly comes from configurations not clustered 
in this way.

As another example, let us look at the system $^9$Be, which supposedly consists
mainly of a two-alpha cluster configuration forming $^{8}$Be with an 
additional neutron to bind the system. This nucleus and the reactions 
$^9$Be$+\gamma \rightarrow \alpha+ \alpha +n$ and its inverse are
of interest for bridging the $A=8$ gap in some astrophysical scenarios,
e.g., in the high-entropy bubble in type II supernovae \cite{EOPT01}.
For the integrated cross section $^9$Be $\rightarrow$ 2 $\alpha$ + n
we find 
\begin{eqnarray}
&&\int d\omega \sigma(\mbox{$^9$Be $\rightarrow$ 2 $\alpha$ + n })
\approx
\int d\omega \left[\sigma(^9\mbox{Be}) - 2 \sigma(^4\mbox{He}) - 
\sigma(n)\right]\\
&&= \left(2 \frac{4}{4} - \frac{16}{9}\right) 60 \mbox{MeV mb} = \frac{2}{9}
60 \mbox{MeV mb} = 13.3 \mbox{MeV mb}.
\end{eqnarray}
(which is identical to the cross section difference one would expect going
only through the unstable nucleus $^8$Be alone). In this case the cross
section at astrophysical energies is dominated by a transition $P_{3/2}
\rightarrow S_{1/2}$.
The experimental results at energies below 2.2~MeV (\cite{EOPT01} and 
references therein) only give about 0.33~MeV mb, that is a very small fraction 
of this sum rule. On the other hand it is well known \cite{Russell48,GuthM49}
that the cross section has a minimum at around 2.2~MeV before rising again
due to the $P_{1/2}\rightarrow D_{5/2}$ transition.

On the other hand the cross section difference of more complex
reactions can be studied in the same way. For both $^6$Li and $^{16}$O we can 
give results for the breakup to a $p+n$ final state, related to the 
breakup of the nuclei in the quasideuteron mode, which is a strong channel.
We find a sum rule prediction of $1/3$ of the total integrated photo cross 
section for the reaction $^6\mbox{Li} + \gamma \rightarrow \alpha + p + n$
and $1/9$ for the reaction $^6\mbox{Li} + \gamma \rightarrow ^{3}He + t$.
The integrated cross section of the ``quasideuteron''
reaction $^{16}\mbox{O} + \gamma \rightarrow ^{14}\mbox{N} + p + n$
is $1/8$ of the total integrated cross section, whereas the single neutron
emission $^{16}\mbox{O} + \gamma \rightarrow ^{15}\mbox{N} + n$ only $1/15$.
In this case and also the other following cases
the cross section has contributions at much higher energies in contrast to
the ``soft dipole modes'' studied here up to now. 
A comparison with experiments, which could be done for both the sum rule
of Eq.~(\ref{eq:ggtsr}) and the one from the spectator approximation 
Eq.~(\ref{eq:ggtsr2}) is difficult in these cases, 
as only a few results are available in the literature \cite{Ajzenberg88}
and most of them also not for these exclusive channel.

Finally in \cite{Grigorenko01} experimental and theoretical problems with
2p-radioactivity and three-body decay are discussed. They also mention 
a number of candidates for some genuine three-body decay. One could again
apply the GGT cluster sum rule to the three-body decay in photoexcitation
experiments. Most of the time one expects the contribution to the cross section
integral to come not only from lower energies. But if a single resonance
dominates the decay, one could again study the question whether this resonance
saturates a major part of the cluster sum rule.

\section{Conclusions}
An alternative derivation of the so-called ``cluster sum rule'' was given,
based on the GGT instead of the TRK sum rule. Whereas the final result is 
formally identical to the ``usual'' cluster sum-rule, the assumptions to
derive this sum rule are different and the 
deviations of the measured from the theoretical results are expected to be 
smaller than in the case of the ``usual'' GGT sum rule. 
Whereas the sum rule of Eq.~(\ref{eq:ggtsr}), expressed as the difference of
two different total photoabsorption cross sections, is quite general, 
we have also looked at the case of a pronounced clustering structure, where
one can make use of the ``spectator approximation'' to identify this 
difference as the cross section for the nucleus going into a certain
final state, see Eq.~(\ref{eq:ggtsr2}). This can be seen as a test of 
the degree to which the clustered configuration contributes to this cross 
section
channel. We have looked at applications of this sum rule for neutron-halo 
systems, as well as, for clustered nuclei and found some agreement with 
experimental findings. We think that this derivation gives
some independent insight into the structure of exotic nuclei but also the 
foundations of the cluster sum rule for exotic nuclei.

We know from many cluster conferences \cite{ClusterX} that the concept of a 
cluster is really quite a loose, qualitative concept. We know much more 
about nuclei from microscopic approaches, like the shell model. However, the 
cluster sum rules discussed in this paper give us a (semi)quantitative measure
of the importance of certain cluster degrees of freedom. We may assume that 
such degrees of freedom become more and more relevant when going away from the
valley of stability, which was found to lead in some cases to a decoupling of 
some nucleons (neutrons) from
the rest of the system. One evidence of this is the well established low lying
dipole strength ({\em ``pigmy resonance''}) in nuclei away from the valley of 
stability.   

\section*{Acknowledgements}
The interest of one the authors (K.H.) in sum rules for nuclei goes back to 
the Les Houches summer school 1996 ``Trends in nuclear
physics, 100 years after'' \cite{LesHouches}. The considerations in the 
present paper grew out of the work on a review article on Coulomb excitation 
for nuclear structure and astrophysics \cite{BaurHT03}.

\end{document}